\documentclass[11pt,a4paper,notitlepage]{article}
\usepackage{geometry}             
\usepackage{extsizes}
\usepackage{amssymb}
\usepackage{amsmath}
\usepackage{amsthm}
\usepackage{subfigure}
\usepackage{multirow}
\usepackage[lined,algonl,boxed]{algorithm2e}

\usepackage{fancyhdr}
\newif\ifpdf
\ifx\pdfoutput\undefined
   \pdffalse
\else
   \pdfoutput=1
   \pdftrue
\fi

\ifpdf
   \usepackage[pdftex]{graphicx}
\else
   \usepackage{graphicx}
\fi

\usepackage{epsfig}

\newcommand{\E}{\mathop{\mathbb{E}}{}\!}

\newcommand{\Eab}{\mathop{\mathbb{E}}{^{\alpha,\beta}}\!}

\newcommand{\Q}{\mathop{\mathbb{Q}}{}\!}
\newcommand{\F}{\mathop{\mathcal{F}}{}\!}

\newcommand\Aab{\ensuremath{A_{\alpha,\beta}}}
\newcommand\Rab{\ensuremath{R_{\alpha,\beta}}}
\newcommand\Ta{\ensuremath{T_{\alpha}}}
\newcommand\Tb{\ensuremath{T_{\beta}}}
\newcommand\Ti{\ensuremath{T_{i}}}

\newcommand\be{$$}
\newcommand\ee{$$}
\newcommand\ben{\begin{equation}}
\newcommand\een{\end{equation}}
\newcommand\bea{\begin{eqnarray*}}
\newcommand\eea{\end{eqnarray*}}
\newcommand\bean{\begin{eqnarray}}
\newcommand\eean{\end{eqnarray}}

\newcommand\Sab{\ensuremath{S_{\alpha,\beta}}}
\newcommand\Survive{\ensuremath{{\rm Survive}}}

\begin{document}

\author{Chris Kenyon\footnote{Affiliation, chris.kenyon@yahoo.com}}
\title{{Completing CVA and Liquidity: \\ Firm-Level Positions and Collateralized Trades}
\footnote{{\bf The views expressed are those of the author only, no other representation should be attributed.}\hfil\break \hbox to 10pt{} The author wishes to aknowledge useful discussions with Roland Stamm and Shane Hughes.}}
\date{16 September 2010, Version 1.01}

\maketitle

\begin{abstract}

Bilateral CVA as currently implement has the counterintuitive effect of profiting from one's own widening CDS spreads, i.e. increased risk of default, in practice\footnote{See for example the \$2.5B profit reported for bilateral CVA, explicitly from own-CDS spread widening, in www.citigroup.com/citi/press/2009/090417a.htm}.  The unified picture of CVA and liquidity introduced by Morini \&\ Prampolini 2010 has contributed to understanding this.  However, there are two significant omissions for practical implementation that come from the same source, i.e. positions not booked in usual position-keeping systems.  The first omission is firm-level positions that change value upon firm default.  An example is Goodwill which is a line item on balance sheets and typically written down to zero on default.  Another example would be firm Equity.  The second omission relates to collateralized positions.  When these positions are out of the money in future, which has a positive probability, they will require funding that cannot be secured using the position itself.  These contingent future funding positions are usually not booked in any position-keeping system.  We show here how to include these two types of positions and thus help to complete the unified picture of CVA and liquidity.   

For a particular large complex financial institution that profited \$2.5B from spread widening we show that including Goodwill would have resulted in a  \$4B loss under conservative assumptions.  Whilst we cannot make a similar assessment for its collateralized derivative portfolio we calculate both the funding costs and the CVA from own default for a range of swaps and find that CVA was a positive contribution in the examples.

\end{abstract}

\section{Introduction}

Recently bilateral counterparty valuation adjustment (CVA) \cite{BP09,BM05,BC10} has been introduced by a number of large complex financial institutions (LCFIs) leading to positive pnl effects at the firm level with increasing probability of self-default.  Additionally, \cite{MP10,Pit10} have highlighted the significance of funding costs for asset pricing by including the effect of funding, aka liquidity, costs into the pricing of individual assets.  Here we point out that for correct calculation of bilateral CVA additional trades must be included that are not booked in any system.  We use two examples to make our point concrete: costs of self-default; and funding costs of unfunded collateralized trades.  At least the second of the examples may appear paradoxical but we demonstrate that in both cases trades that are not booked are significant.  This contribution goes some way to resolving the paradox of benefiting from your own default and completing the unified picture of CVA and liquidity.

At the firm level we consider Goodwill as an example of an asset that is usually valued at zero after default.  Whilst experts in Corporate Finance will be aware that the valuation of Goodwill is highly subjective \cite{RW10}, it is a line item on corporate balance sheets and reported regularly.  Other examples of such firm-level assets also exist, e.g. Equity.  This is also a line item on balance sheets and the majority of this is written down on default.  Since bilateral CVA has been used to profit from spread widening on booked trades at the firm level, the debate on which corporate balance sheet items should also be included in bilateral CVA is somewhat overdue.  This paper makes a quantitative start on the debate. 

When a collateralized trade is out of the money collateral (often cash) must be posted, and hence funded.  We quantify these funding costs for a swap in isolation.  The implications at the firm level derive from netting and the overall funding position --- but without a knowledge of future funding requirements and costs the picture is incomplete.  Collateralized trades have been considered in CVA calculations \cite{Greg10} --- but not in combination with their future contingent funding requirements.  In as much as future funding is required then these funding trades contribute to bilateral CVA.  If net funding requirements oscillated daily about zero then it would make sense to fund daily.  However, banks typically do require significant short-term funding as the crisis has made clear (this does not separate asset funding form derivative funding).  From a theoretical point of view, random walks typically spend long periods away from their origin\footnote{A simple random walk on a $d$-dimensional lattice is recurrent only for $d=1,2$ (P\'olya's Theorem, 1928).} which would also imply the accumulation of net funding positions.  Again, in as much as funding is significant this will be funded at longer terms --- precisely to avoid market or operational disruptions.  Finally, capital market funding (i.e. short term) is used when banks surf the yield curve, i.e. borrow short and lend long.

Collateral funding is not a zero sum game, at best an institution breaks even with zero net funding.  For all other funding cases funding costs.  We use our vanilla collateralized swap as a starting example for quantifying funding costs and their CVA for collateralized trades.  In as much as future contingent funding trades are not already booked then this is another gap in the current implementation of bilateral CVA.

Thus by considering trades and position that are not present in typical position-keeping systems, i.e. unbooked trades, we help to complete the picture of CVA, especially at the level of the firm.

\section{Unbooked Positions} 

\subsection{Firm-Level Assets and Liabilities}

CVA is typically calculated for trades and assets in position-keeping systems.  However, significant assets and liabilities are not booked on these systems.  This means that CVA is incomplete with respect to firm-level assets and liabilities.  To illustrate the need for completeness, i.e. including assets that are not in the position-keeping systems, we consider Goodwill.  Goodwill is a controversial and subjective topic in the corporate finance world.  However, in as much as it is a line item on a balance sheet it is a concrete asset.  Goodwill is listed on financial reports that must be kept up to date by the firms that report it.  On default of the firm Goodwill is usually reduced to zero --- indeed this can happen even on near default when customers lose confidence in a firm, or when the firm's brand value is lost.

Goodwill is not the only firm-level asset we could consider, for example a firm's Equity is usually damaged by default as well.  We work with Goodwill as our example because it fits with what we want to consider: it is a firm-level asset (and a line item on a firm's balance sheet); it is not booked in position-keeping systems; and it is usually valueless on firm default.  However, we note that other firm-level assets, such as Equity, should be included in any firm-level reporting.

Typically bilateral effects at the trade level are added up and presented as firm-level effects, leading to the phenomenon of benefitting from one's own default.  For consistency, if bilateral CVA is used in firm-level accounting then it should also be applied to firm-level assets, for example Goodwill, going-concern value etc.  The basic idea of bilateral CVA is to continue valuation past the default event.  This event has implications beyond just the trades booked in position keeping systems.  We can include these unbooked assets within bilateral CVA to help complete the PnL picture.  We show that including these firm-level assets in the firm-level picture significantly changes the picture of benefitting from one's own default into something much more intuitively reasonable.

We use the example of Goodwill as an unbooked asset (unbooked at least in position-keeping systems).  In some cases Goodwill may be difficult to assign calculate, however in the case of takeovers Goodwill is generally a clearly defined value.  In addition Goodwill is a line item on quarterly corporate reports.   

We consider a large complex financial institution (LCFI) example from 2009 since it made the headlines for its use of CVA to profit from its increased CDS spreads, i.e. likelihood of default.  The quote below is from the LCFI's first quarter 2009 earnings release:
\begin{quote}
$>$ A net \$2.5 billion positive CVA on derivative positions, excluding monolines, mainly due to the widening of XXX's CDS spreads.\footnote{www.citigroup.com/citi/press/2009/090417a.htm}
\end{quote}
We note from the LCFI annual reports from 2008 and 2009 that Goodwill was \$27B at the end of 2008 and \$26B at the end of the first quarter 2009.  There was no mention of negative CVA on Goodwill.  Between the end of 2008 and the end of Q1 2009 the LCFI's 5 year CDS spread moved from roughly 196bps to roughly 667bps, at the same time the LCFI's 20 year CDS spread moved from roughly 374bps and 532bps.

The fair value of Goodwill must be updated by companies from time to time according to accounting standards\footnote{US GAAP FAS 142, IFRS 3.  N.B. the author gives no guarantees regarding the present interpretation: consult an appropriate professional for financial, accounting or legal questions.}, rather than amortized over time.  Depending on the model of future value of Goodwill a wide range of outcomes are possible.  We provide a range of alternatives since there is no consensus w.r.t. CVA as we are introducing this concept (applying CVA to Goodwill) in the present paper.  This illustrates the uncertainties with moving CVA to the level of the firm.  We consider the following models for the development of Goodwill value.
\begin{description}
\item[AMORTIZING] the updated value of Goodwill does, in fact, decrease linearly over some fixed horizon.
\item[CONSTANT] the Goodwill has a constant value.  Thus a Goodwill value observed today as \$10M will remain \$10M perpetually until default.  This is the limit of AMORTIZING with an infinite horizon.
\item[STOCK] Goodwill has a constant expected discounted risk neutral value.  This effectively models Goodwill like a stock, i.e. the risk neutral future value of the Goodwill is the same as its present value (i.e. it is a martingale).
\end{description}

The CVA for Goodwill depends critically on the future development of its value, the more it is worth later the higher the CVA.  At the limit, we can expect any company to default at some point even if this is, for example, 100 years in the future (the maturity of the longest GBP swap on Bloomberg is 70 years).  If Goodwill is a martingale under the risk neutral measure then its CVA is 100\%\ of its current value and this value is independent of the CDS spread of the firm.  Alternatively we can look at the company's survival probability and choose some fraction of its expected life (or time to survival probability of some percentage).

\be
{\rm CVA}_{GW} (t,T)= \E_t\left [\int_{s=t}^{s=T} G(s) df(t,s)  \lambda(s) e^{-\int_{u=t}^{u=s} \lambda(u) du} ds | \F_t \right]
\ee
where:

$T$ horizon for Goodwill CVA calculation,

$G(s)$ Goodwill value at time $s$,

$\lambda(s)$ hazard rate,

$df(t,s)$ discount factor from $t$ to $s$,

\begin{figure}[htb]
\begin{center}
\includegraphics[width=0.8\textwidth,  clip=true, trim=10 0 0 40 ]{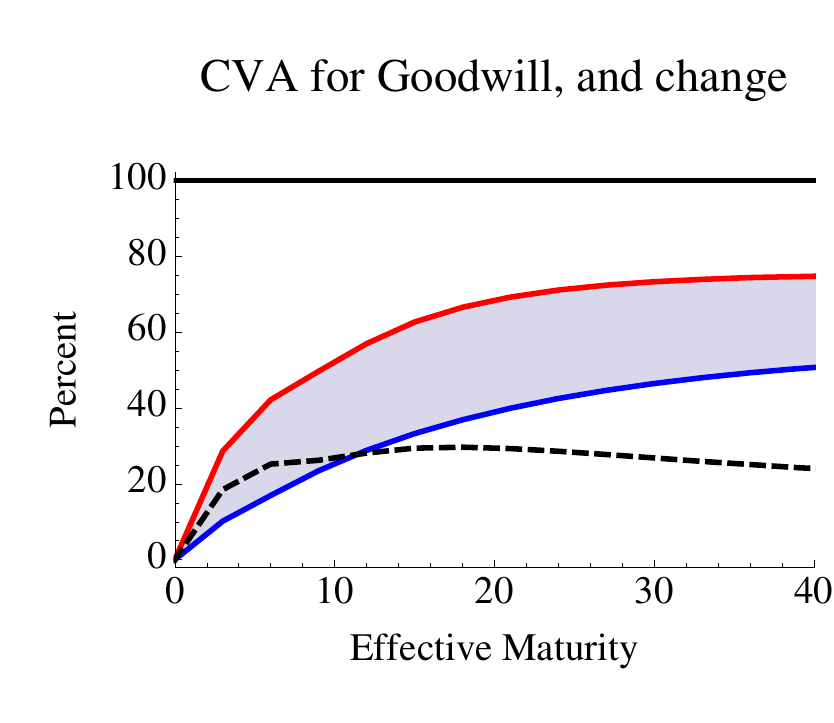}
\caption{CVA for Goodwill for a particular LCFI versus its effective maturity in years assuming linear amortization in practice (not by regulation) asof 2008YE (lowest full line), asof 2009Q1 (middle full line).  The shaded area represents the change from 2008YE to 2009Q1.  The dashed line gives the change in CVA for Goodwill from 2008YE to 2009Q1.  This change is roughly 20\%\ to 25\%.   The top, constant, line shows the CVA for Goodwill at any date assuming that the discounted Goodwill value is a martingale in the risk neutral measure.\label{fig:goodwill}}
\end{center}
\end{figure}

\begin{figure}[htb]
\begin{center}
\begin{tabular}{| l | c | c | c | c | c | c | c | c | c | c |} \hline
maturity & 0.5 & 1 & 2 & 3 &  4 & 5 & 7 & 10 & 15 & 20 \\ \hline
2008YE & 262 & 262 & 230 & 218 & 203 & 196 & 196 & 196 & 196 & 196 \\ 
2009Q1 & 923 & 923 & 800 & 701 & 665 & 638 & 581 & 534 & 534 & 534  \\ \hline
\end{tabular}
\caption{CDS spreads (bps) used for example for the LCFI for year end 2008 and end of first quarter 2009. Bloomberg (shown here) and Markit give similar numbers, except for the last point at 20Y where Markit shows a very significant increase.  Recovery rate was taken conventionally as 40\%. }
\end{center}
\end{figure}

\vskip5mm

Considering the LCFI example it appears that the LCFI do not model future Goodwill value as STOCK because in this case the CVA adjustment would be 100\%, i.e. \$27$\sim$\$26B or  on both sets of accounts.  If the LCFI picked AMORTIZING then we must also decide the amortization maturity.   We observe from Figure \ref{fig:goodwill} that the Goodwill changed between 2008YE and 2009Q1 by 20\%\ to 25\%\ depending on the effective amortization maturity chosen.  In dollars this is 25\%\ of $\sim$\$26B between 2008YE and 2009Q1.  This would have changed the sign for the CVA:  the CVA goes from a benefit of ${}+{}$\$2.5B to a CVA loss of about \$4B\footnote{Using Markit data (with the CDS spread increase for the Y20 point) we observed two regimes for the CVA change result: maturity of less than 17 years; maturity of greater than 17 years.  Twenty years corresponds roughly to the point at which the survival probability of the LCFI was 20\%\ for both dates.  If we consider a conservative effective amortization maturity of less than 17 years then the change in their CDS curve would indicate that the CVA adjustment would have changed by 25\%$\sim$30\%\ of $\sim$\$26B between 2008YE and 2009Q1.  This would have changed the CVA from a benefit of \$2.5B to a loss of $\sim$${}{}$\$4B.  That is, substantially the same result with Markit data as with Bloomberg data.}

Note that there is a significant CVA for the Goodwill in all future Goodwill models --- the higher the future Goodwill the bigger its CVA.

\subsection{Contingent Funding for Collateralized Positions}

Collateralized positions are unbooked to the degree that their future funding requirements are not captured in position-keeping systems.  Consider a standard swap contract, at inception its NPV is zero.  On trade date ATM swaps require no funding.  CVA applied to uncollateralized swaps recognizes that at later times they may be in the money and thus they require a counterparty adjustment.  For collateralized swaps, the swap NPV itself requires no counter-party adjustment (assuming no gap risk, and that collateral calls are made daily, and ignoring one-day moves).  In reality, disputes and other delays create a period of several days in which the called-on value of the derivative is at risk.  \cite{Greg10} assumes 10 days, so the usual VaR methodology applies.  However, we concentrate here on the funding aspects.

Now, when the swap is out of the money in the future this must be funded.  CVA for collateralized products has been considered in the literature \cite{Greg10} (Chapter 5) but not including funding, and funding is the critical point here since the funding trades cannot (by definition) be collateralized themselves.  Typically, banks do not book contingent funding trades for potential future collateral requirements.  Thus the funding must come from unsecured funding as the standard swap in our example is not an asset that can be placed into repo.  We can make a first estimate of this cost for a standard swap by modeling its future value on coupon dates, i.e. assuming that the collateral calls are on those dates and ignoring market movements between them.  Of course the market will move between coupon dates so this is a first approximation.

\subsection{Single Swap}

We consider two banks, Y and Z.  A fair vanilla fixed-for-floating swap has value zero on trade date.  At later dates it has non-zero value, and the price of co-terminal swaptions for a strike equal to the fixed rate in the fair swap give the risk neutral information on how the market sees the future value of its remaining length asof today.  

When a swap is collateralized bank Y receives collateral, here assumed to be cash, when it is in the money.  Bank Y posts collateral when it is out of the money (OTM).  This cash must be funded unsecured w.r.t. the swap considered in isolation.  The swap itself cannot be used as collateral when OTM since it is OTM.

\subsubsection{Calculations}

We can model the cost of the collateral postings by both banks at time zero assuming funding trades at interval $\tau$:
\bean
K= S_{0,\beta}(0) && \nonumber \\
{\rm SWAP_{FUNDING}}(0) &=& \sum_{\alpha=\tau}^{\alpha=\beta-\tau}  \Aab(0) \Eab
\left[ \tau (F_{\tau}(T_{\alpha}) - FOIS_{\tau}(T_{\alpha})) \right. \nonumber \\
&&\ \ \ \ \ \ \ \ \  \times \left.    \left( \Sab(T_{\alpha}) -  K \right)^+ \right] \label{eq:funding}
\eean
where:

$\Aab(t)$ is the value of the Annuity associated with the swap \Sab,

$\Eab$ is the expectation relative to the Annuity measure.

$F_{\tau}(T_{\alpha}) $ is the tenor $\tau$ forward rate at $T_{\alpha}$ (N.B. in Annuity measure)

$FOIS(T_{\alpha})$ is the tenor $\tau$ forward rate based on the overnight index at $T_{\alpha}$ (N.B. in Annuity measure)

$ \Sab(T_{\alpha})$ fair swap rate at $T_{\alpha}$.

\vskip3mm

Above we assume that both banks fund at Libor flat and receive OIS (overnight rate) from posted collateral. Note that the banks cannot fund at OIS for the collateralized trade itself because it is not available as security to fund against.  Basically the banks are paying unsecured and receiving overnight when they have to post cash as collateral.  

If a bank is paying more than Libor flat for funding we must include this in the equation above, modifying $F_{\tau}(T_{\alpha})$.  It might be argued that a bank can fund overnight and so only pay OIS for funding.  Firstly this is only true if the funding is secured.  Secondly, no bank willingly gets its funding overnight because then the slightest operational or market disruption would produce immediate issues.  It is normal to roll capital market funding at some tenor (e.g. 3 months or 6 months) for the vast majority of the amount.  Funding trades will be done every day, just not for a tenor of one day.  Some minor day-to-day changes will be done using overnight, but only a very small proportion of the total.

We approximate Equation \ref{eq:funding} as (ignoring for the moment the change of measure for the forward rate):
\bean
{\rm SWAP_{FUNDING}}  &\approx & \sum_{\alpha=\tau}^{\alpha=\beta-\tau} \tau 
 \Aab(0) \Eab \left[   (F_{\tau}(T_{\alpha}) - FOIS_{\tau}(T_{\alpha})) \right]   \label{eq:fundingapprox1}\\
&&\ \ \ \ \ \ \ \ \  \times   \Eab \left[ \left( \Sab(T_{\alpha}) -  K \right)^+ \right]  \nonumber  \\
&\approx & \sum_{\alpha=\tau}^{\alpha=\beta-\tau} \tau 
 \Aab(0) \Eab \left[   FOO_{\tau}(T_{\alpha},T_{\alpha}) ) \right]   \label{eq:fundingapprox2}\\
&&\ \ \ \ \ \ \ \ \  \times   \Eab \left[ \left( \Sab(T_{\alpha}) -  K \right)^+ \right]  \nonumber  \\
&\approx & \sum_{\alpha=\tau}^{\alpha=\beta-\tau} \tau 
 \Aab(0) FOO_{\tau}(0,T_{\alpha})    \label{eq:fundingapprox3}\\
&&\ \ \ \ \ \ \ \ \  \times   \Eab \left[ \left( \Sab(T_{\alpha}) -  K \right)^+ \right]  \nonumber  
\eean
where:

$FOO_{\tau}(U,V)$ is the funding cost for tenor $\tau$ over overnight at time $V$ as seen from time $U$.

\vskip3mm

The  main element to the approximation in Equation \ref{eq:fundingapprox1} is the covariance between the funding cost $F_{\tau}(T_{\alpha}) - FOIS_{\tau}(T_{\alpha})$, actually funding over overnight $FOO$, and the positive part of the difference in future swap prices $()^+$.  Whilst the covariance between forward and swap rates will general be significant for short swaps the covariance between the forward rates and the positive difference will be much lower (because the negative part of the difference will detract from any correlation).  A secondary element to the approximation in Equation \ref{eq:fundingapprox2} is the use, by implication, of a zero correlation between the funding over overnight and the forward rate.  In as much as the included forward OIS and forward CDS rates are not correlated with the standard Forward rates this will be increasingly accurate.  Finally in Equation \ref{eq:fundingapprox3}  we have again assumed a zero correlation between the funding over overnight and the forward rate.

We use the right hand side of Equation \ref{eq:fundingapprox3} as an approximation going forward for calculations.  This means that we can use the swaption cube market data directly, i.e. include the swaption smile easily.  Note that this approximation is conservative and will be more accurate the more the correlation is low, which will be the case for longer swaps where the funding contributions are higher.

The CVA from own default w.r.t. this funding cost is, assuming independence of (forward) CDS spread volatility and interest rate volatility, or equivalently deterministic hazard rates:
\bean
{\rm SWAP_{CVA}}(0) &=& LGD \sum_{\alpha=\tau}^{\alpha=\beta-\tau}  
\left( \Survive(T_{\alpha}) - \Survive(T_{\alpha}+\alpha)\right) \label{eq:swapcva} \\
&& {}\times 
\Aab(0) \Eab \left[ \left( \Sab(T_{\alpha},{\rm ATM}) -  K\right)^+ \right]    \nonumber
\eean
where:

$LGD$ is the loss given default of the LCFI;

\Survive(S)\ is the survival probability to time $S$.

\vskip3mm

The logic is that the funding is only taken on providing the bank survives to the start of a funding period, and the funding is only not paid back if default occurs before the funding is rolled.  

Note that there is no CVA for counterparty default --- the collateral takes care of this (up to gap risk and overnight moves, see \cite{Greg10} for more details).

\subsubsection{Funding Costs\label{ss:fund}}

\begin{figure}[htb]
\begin{center}
\includegraphics[width=0.8\textwidth,  clip=true, trim=0 0 0 24 ]{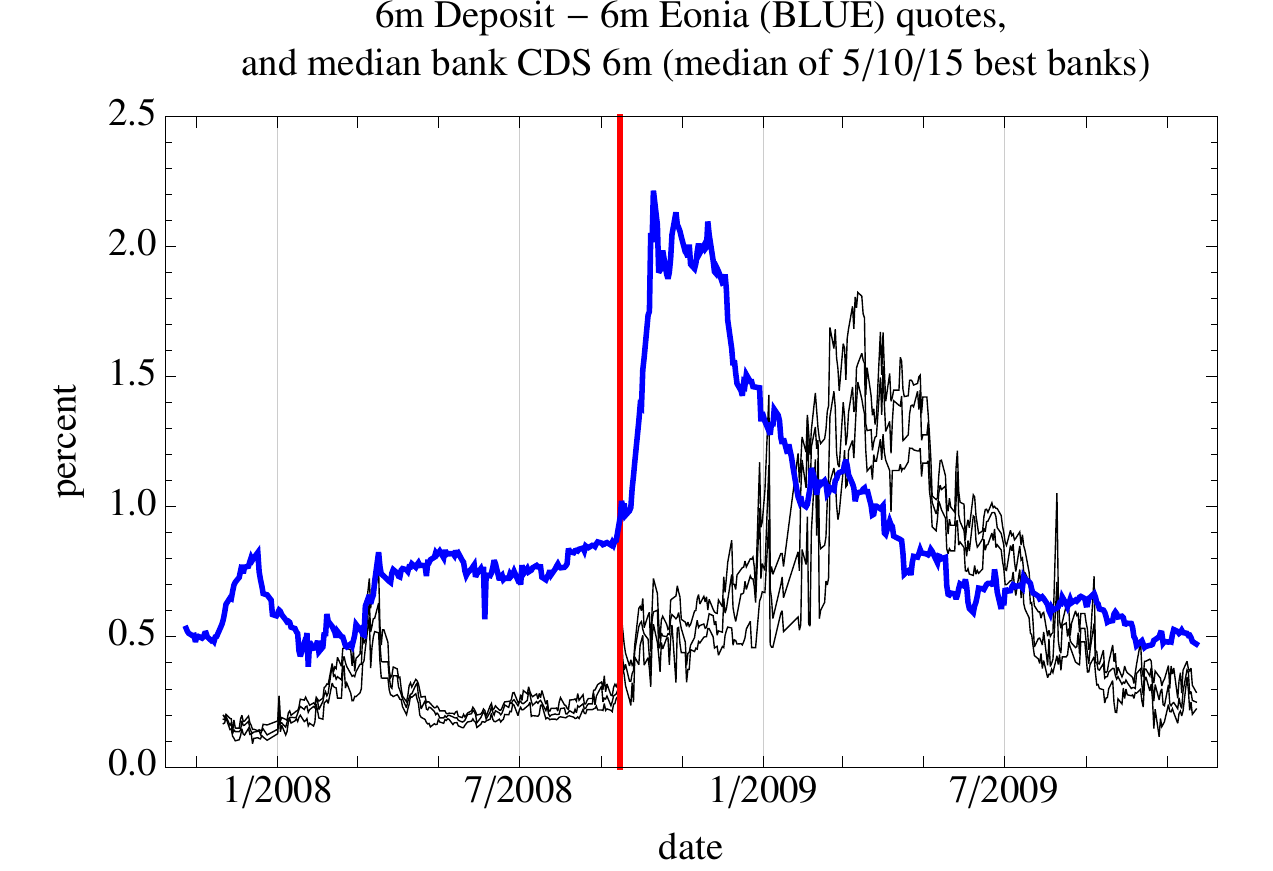}
\caption{Components to funding costs.  Vertical line is on the date of Lehman default.  Upper thick line is 6 month Deposit rate minus 6 month Eonia.  Lower thin lines are the medians of the CDS spreads of the best 5/10/15 A-rated banks with Euro-quoted CDS spreads.  (N.B. the 6 month Euribor rate is very close to the 6 month Deposit rate).\label{fig:fundingparts}}
\end{center}
\end{figure}
\begin{figure}[htb]
\begin{center}
\includegraphics[width=0.8\textwidth,  clip=true, trim=0 0 0 10 ]{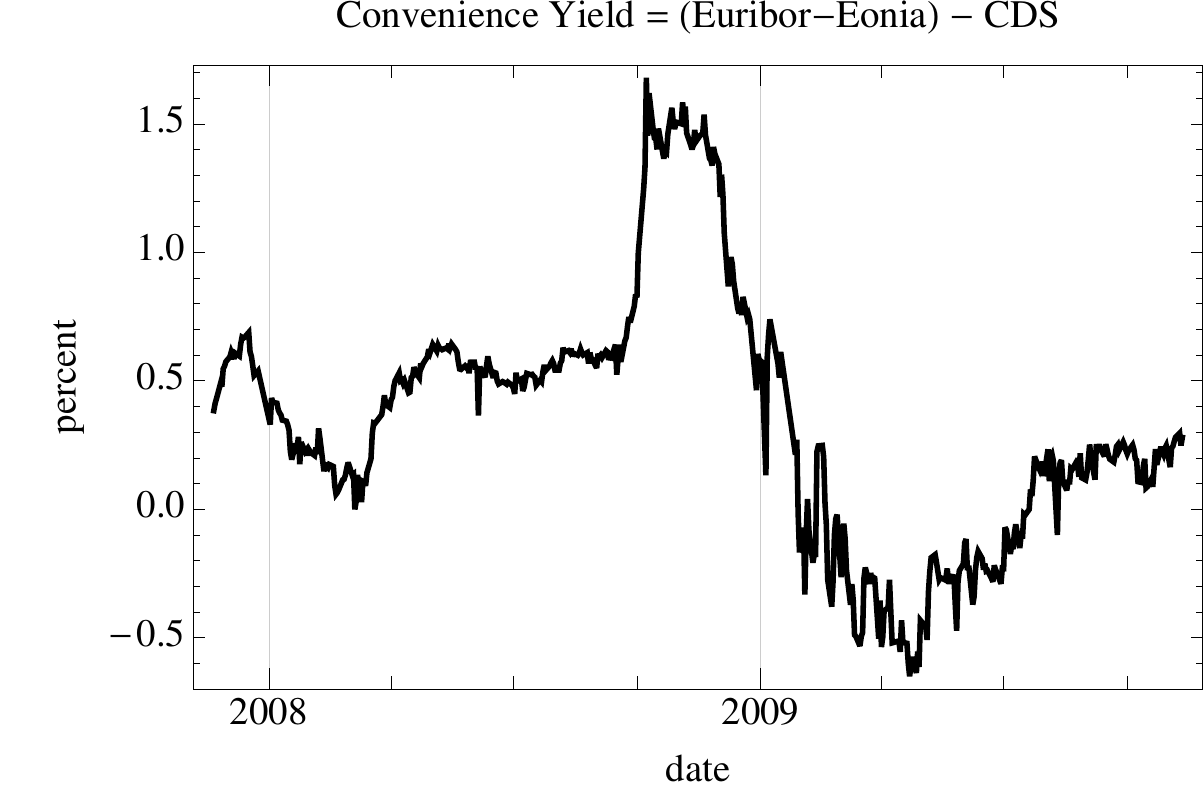}
\caption{Cash scarcity spread as (6 month Deposit rate minus 6 month Eonia) $-$ median CDS spread of best 10 A-rated banks with Euro-quoted CDS spreads.  This is negative from end January 2009 through July 2009, note however that funding costs will still be positive.  \label{fig:convenience}}
\end{center}
\end{figure}

We need a funding cost to calculate CVA for collateralized trades which are funded from the capital markets when OTM (i.e. posting collateral so having to raise it).  Since our example LCFI above explicitly referenced CDS spreads we need to distinguish the non-CDS component of funding costs.  We do this very simply by assuming (for both spot and forward):
\ben
\rm funding \ spread\ over \ overnight = credit\ spread + scarcity\ spread \label{eq:fundingparts}
\een
In the commodity literature a scarcity spread is also referred to as a convenience yield.  When negative it is typically referred to as a storage cost (or yield).  In general it can be positive or negative although it is usually positive.  We assume that the scarcity spread is a market-wide phenomenon whereas the credit spread is specific to a particular institution.  Thus we can estimate the market-wide scarcity spread directly as the difference between the left hand side of Equation \ref{eq:fundingparts} and an appropriate credit spread.  We assume that the appropriate credit spread is the average CDS spread of a set of good banks, e.g. the banks that make up the Eonia/Euribor committee.   

For this paper we use the median spread of the best 10 banks in the Eurozone with at least an A rating to calculate the market-wide (spot) scarcity cost, see Figure \ref{fig:fundingparts}.  By "best banks" we also mean that they had CDS spreads of less than 50bps pre-crisis in January 2008.
 Note that in Figure  \ref{fig:fundingparts} the scarcity cost jumps from about 50bps pre-Lehman to about 150bps just post-Lehman.  The next phase of market development is the central banks adding liquidity, so reducing the scarcity spread, and the increase in banks' cds spreads peaking around May of 2009.  At this point, and at the end of 2009 Q1 the scarcity spread is actually negative 50bps.  
 
It might appear from Figure  \ref{fig:fundingparts} that there was an arbitrage opportunity around May 2009 when banks' cds spreads were higher than funding costs.  However, this is not the case because a short position is very different to a long position in terms, precisely, of funding costs.  Thus it is legitimate for the CDS curve to be above Euribor - Deposit.

For forward funding costs we use exactly Equation \ref{eq:fundingparts} now applied to the forward credit spreads and forward funding spreads.  The equation for the forward CDS rate $\Rab(0)$ at time zero is:
\be
\Rab(0) = \frac{LGD \int_{u=\Ta}^{u=\Tb} df(u) d(e^{-\Gamma(u)})  }{  \sum_{i=\alpha+1}^{i=\beta} \tau_i  e^{-\Gamma(\Ti)}  }
\ee
as in \cite{brigo_interest_2006} where:

$df(s)$ is the discount factor to time $s$;

$e^{-\Gamma(s)} = \Q(\kappa > s)$  the survival probability to time $s$ ($\kappa$ is the default time).

\begin{figure}[htb]
\begin{center}
\includegraphics[width=0.6\textwidth,  clip=true, trim=0 0 0 00 ]{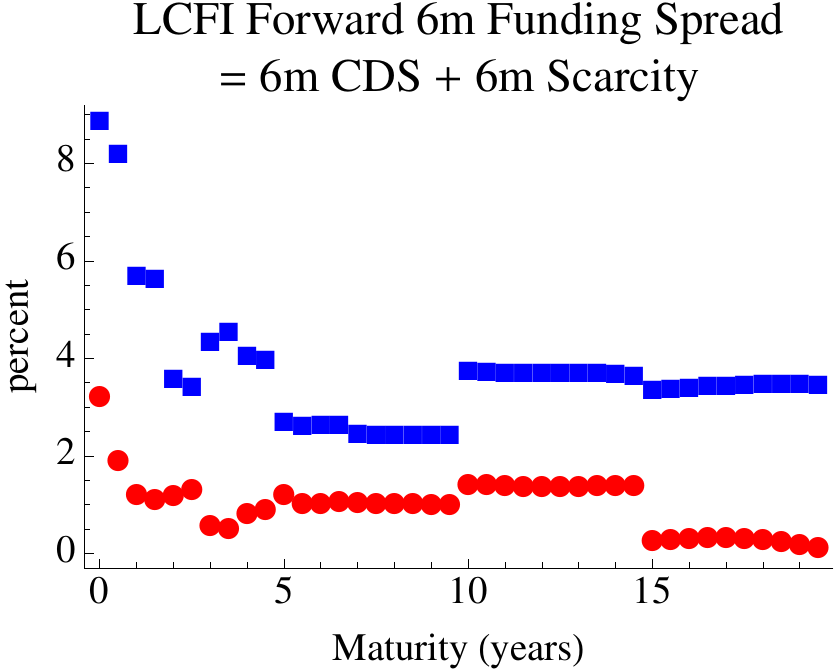}
\caption{Forward funding rate for the LCFI at 2008YE and 2009Q1.  As a conservative estimate of the funding costs for the LCFI we use the average values which are 1.0\%\ and 3.7\%\ respectively for all maturities.  \label{fig:lcfiFunding}}
\end{center}
\end{figure}

\subsubsection{Example}

For our example we consider swaps traded by the large complex financial institution (LCFI) mentioned above on 2008 year end and 2009 quarter 1.  We need overnight and tenor-based curves for building forward rates.  Since it is easier to obtain long (i.e. 20 year) maturities for both types of swaps in Euro-land we will switch our example from USD to EUR.  This example is illustrative --- we are not implying anything about an actual LCFI although we take inspiration from that context.

It might be argued that a LCFI can fund overnight and so will only pay OIS/Eonia/Sonia for funding.  Firstly this is only true if the funding is secured.  Secondly, no bank willingly gets its funding overnight because then the slightest operational or market disruption would produce immediate issues.  It is normal to roll capital market funding at some tenor (e.g. 3 months or 6 months) for the vast majority of the amount.  Funding trades will be done every day, just not for a tenor of one day.  Some minor day-to-day changes will be done using overnight, but only a very small proportion of the total.

\begin{figure}[htb]
\begin{center}
\includegraphics[width=0.99\textwidth,  clip=true, trim=0 0 0 0 ]{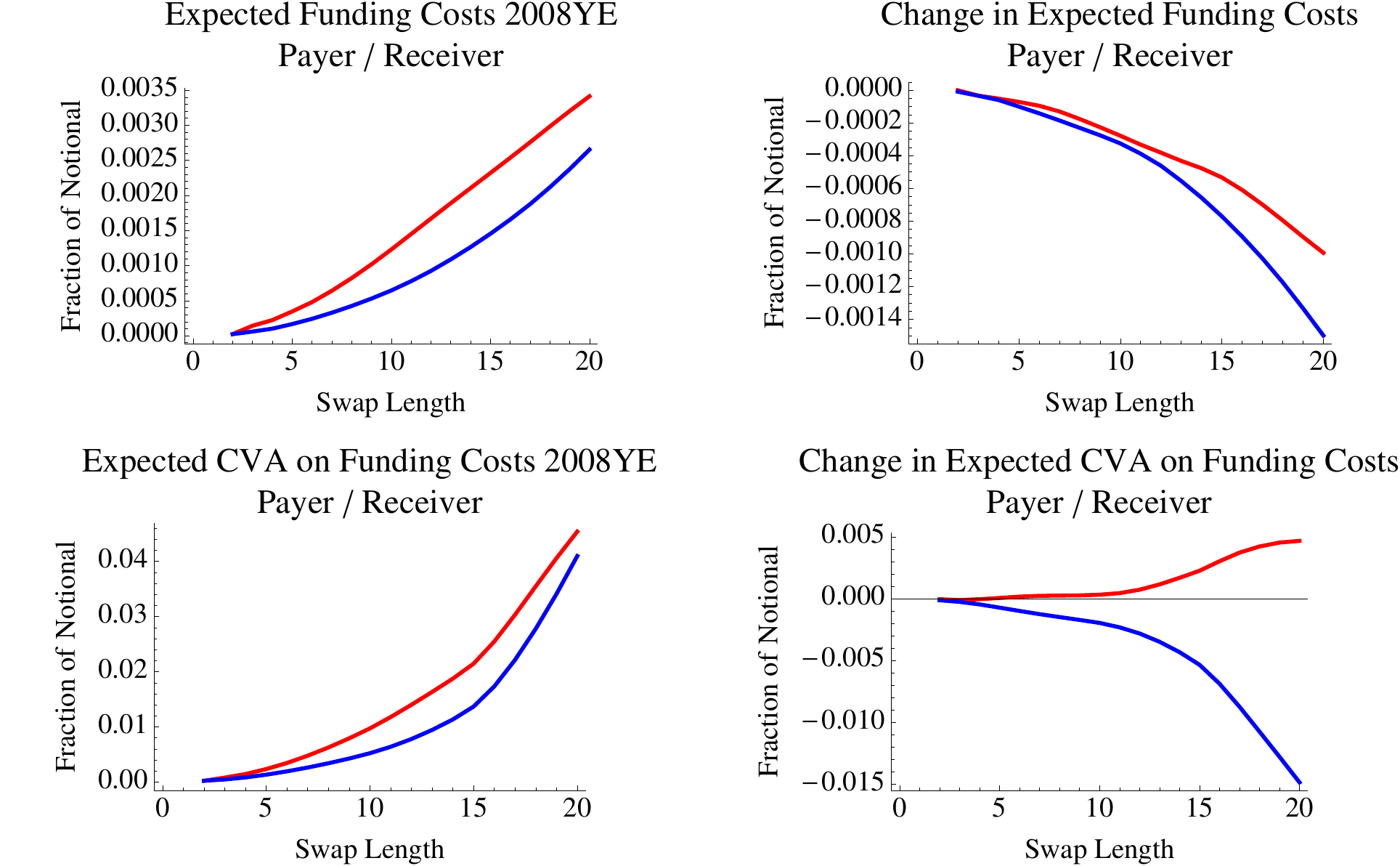}
\caption{Funding at {\bf Libor flat}.
{\it upper: }Payer and Receiver Swap funding costs for spot ATM swaps of different lengths at 2008YE.  
{\it right: }Changes in Payer and Receiver Swap funding costs for spot ATM swaps of different lengths at between 2008YE and 200Q1.
{\it lower:} CVA due to self default on Payer and Receiver Swap funding costs for spot ATM swaps of different lengths at 2008YE.
 {\it right:} Changes in CVA due to self default on Payer and Receiver Swap funding costs for spot ATM swaps of different lengths between 2008YE and 2009Q1.
 \label{fig:flat}
}
\end{center}
\end{figure}

\begin{figure}[htb]
\begin{center}
\includegraphics[width=0.99\textwidth,  clip=true, trim=0 0 0 0 ]{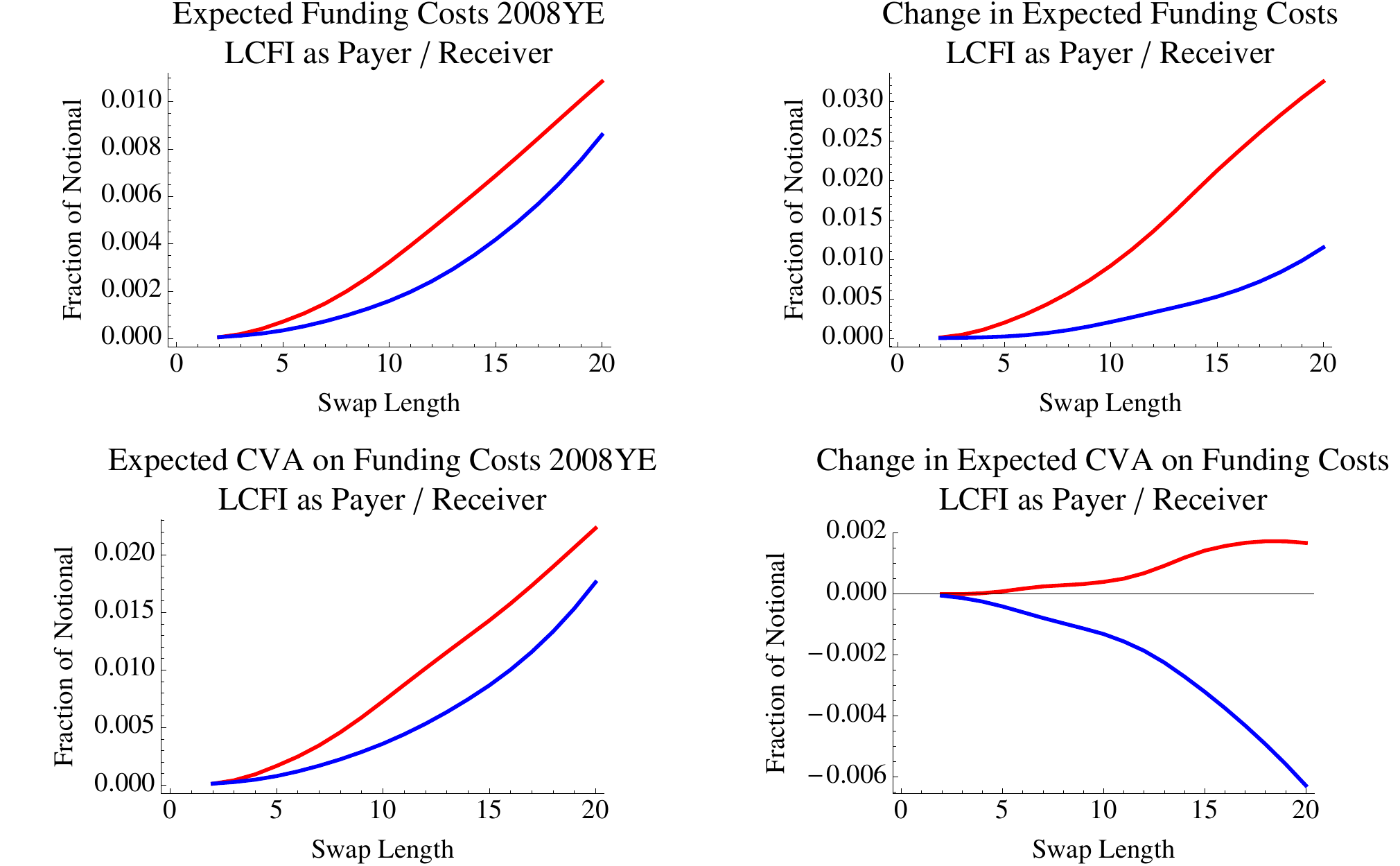}
\caption{Funding at {\bf LCFI level above Eonia}, i.e. 1.0\%\ for 2008YE and 3.7\%\ for 2009Q1.  Note the altered vertical scales for the upper plots. 
{\it upper: }Payer and Receiver Swap funding costs for spot ATM swaps of different lengths at 2008YE.  
{\it right: }Changes in Payer and Receiver Swap funding costs for spot ATM swaps of different lengths at between 2008YE and 200Q1.
{\it lower:} CVA due to self default on Payer and Receiver Swap funding costs for spot ATM swaps of different lengths at 2008YE.
 {\it right:} Changes in CVA due to self default on Payer and Receiver Swap funding costs for spot ATM swaps of different lengths between 2008YE and 2009Q1.
 \label{fig:lcfi}
}
\end{center}
\end{figure}

As a base case we consider a funding spread of zero (i.e. Euribor flat, rolled semi-annually) in Figure \ref{fig:flat}.  The overnight rate for the cash collateral (Eonia), is received for cash posted to the counterparty.  Thus we have net funding costs of Euribor minus Eonia.  We then consider the LCFI in our previous example and take its funding cost over overnight to be forward 6 month scarcity spread (spot is +150bps 2008YE, then spot is -50bps 2009Q1) plus the its (forward) 6 month CDS spread as appropriate in Figure \ref{fig:lcfi}.  Note that we estimate the forward 6 month scarcity spread by using Equation \ref{eq:fundingparts} for forward rates.

\begin{figure}[htb]
\begin{center}
\includegraphics[width=0.6\textwidth,  clip=true, trim=0 0 0 0 ]{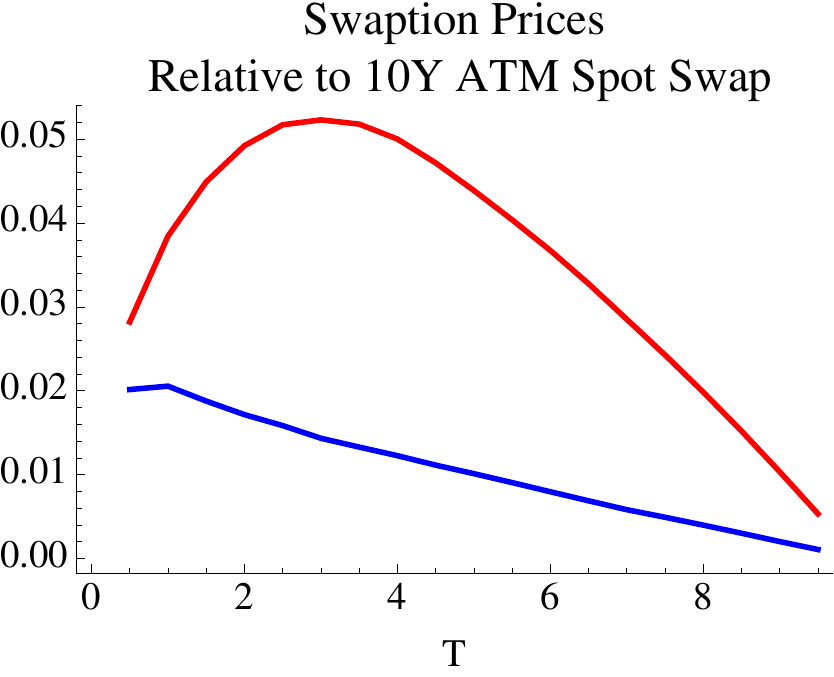}
\caption{Payer and Receiver Swaption prices (lowest and humped lines).  Swaptions are for the remaining length of the swap so although the strike (constant, from 10Y ATM spot swap) is not ATM for these swaps, their value decreases  in the end because the referenced swap length decreases.\label{fig:asym}}
\end{center}
\end{figure}

Note that the funding costs for the two sides (whether long or short) of the swap will not be symmetric --- this is easily understood by looking at Payer/Receiver swaption prices using as strike the ATM swap level (for the original swap), see Figure \ref{fig:asym}.

The resulting expected funding costs for the two counterparty banks are also shown (or alternatively the same bank on either side of a fixed-for-floating).  Note that these are not symmetric.  This is as expected from the fact that the ATM swap rate varies and hence the swaption prices are not equal for payer and receiver.  

\subsubsection{Flat Funding for LCFI}

Figure \ref{fig:flat}. For the vanilla ATM swaps considered, funding costs (assuming funding at Euribor flat for both counterparties, rolled semi-annually and receiving Eonia from posted collateral), were up to 20bps to 35bps (Payer / Receiver) for 20 year maturity.  Note that the funding costs are asymmetric --- hence we expect ATM swaps as market traded to actually be biased.

Bilateral CVA applied to the funding required for the vanilla ATM swaps considered showed a considerable benefit of 150bps to 200bps depending on the side.  The CVA is much larger than the funding cost because on default the whole notional of the funding has a CVA whereas the funding cost is the notional times the difference of two forward rates.  N.B. this CVA benefit is clearly due to the high probability of self default of the LCFI in question.

\subsubsection{LCFI Funding for LCFI}

Figure \ref{fig:lcfi}.  We now consider the same vanilla swaps but with funding costs for the LCFI as calculated in Section \ref{ss:fund}.  The major difference is that the funding costs themselves are much higher (as input to the scenario) at 2008YE.  In addition the increase in funding costs from 2008YE to 2009Q1 is large, up to 300bps for a 20 year swap.  

The CVA on the funding costs is substantially the same with LCFI funding costs as opposed to FLAT in the previous section.  This is because the CVA is dominated by the LCFI default probability which is the same in both funding scenarios.  Since the CVA is substantially the same between the scenarios it is no surprise that the change in CVA is also substantially the same with different funding costs.

\subsubsection{Conclusions on CVA for Collateralized Swaps}

For the example we see that the CVA on the collateralized swaps is substantial as a fraction of swap notional (up to 2\%\ for 20 year swap) and that this changes from 2008YE to 2009Q1 but that the sign of the change depends on whether the swap is fixed-for-floating or floating-for-fixed.  The CVA is almost independent of the funding costs but (as expected) dependent on the default probability. 

\subsection{Swap Portfolios and Netting}

Clearly collateralized positions are netted by counterparty according to CSA (collateral support agreements).  The net posting requires funding and the positions themselves cannot, as mentioned before, be used as security for the funding.  Thus some sort of unsecured funding must be paid.  At the firm level this is generally set internally and charged by funding desks to derivatives desks.

\section{Discussion}

Bilateral CVA calculated using only the trades currently booked and then reported at the firm level leads to intuitively strange effects --- such as profiting from one's own default.  What bilateral CVA does is valuation past the default event of the firm.   We have a number of conclusions for CVA:
\begin{enumerate}
\item  If CVA calculation goes across  the firm default event then firm-level assets that are not booked must also be included for consistency.  In the particular example where a large complex financial institution profited from its CDS spread widening from 2008 Year End to 2009 Quarter 1, including CVA for Goodwill would have reversed the overall effect from a benefit of \$2.5B to a loss of \$4B.  This comes with a CVA for 2008YE of somewhere between 10\%\ and 25\% of the \$27B Goodwill.

Alternatively, with a longer maturity on the Goodwill then the change would have  $+$\$1.5B, but it would have already been 70\%\ written off, i.e. \$18B less. 
\item Collateralized positions require funding, at least in expectation which creates a cost since this cannot be at secured rates.  And the funding costs are asymmetric --- hence we expect ATM swaps as market traded to actually be biased.
\item Bilateral CVA applied to the funding required for the vanilla ATM swaps considered showed a benefit commensurate with the probability of default of the LCFI in question.  However, the CVA was substantially independent of the exact funding level.
\end{enumerate}

We note that the tradeoff for collateralized positions is between the CDS spread and the funding cost.  Unlike the case considered by \cite{MP10} we do not have matched funding, i.e. funding is being done for much shorter times than the life of the derivative.  In fact, since the swap considered is ATM it requires no funding at inception and its funding requirements can change from period to period.

The funding strategy and the portfolio considered here , one swap, are as simple as possible.  However, in as much as a LCFI uses capital market funding, this is typically rolled on a short term basis so our conclusions are relevant.  

With respect to balance sheet items, like Goodwill and Equity, it could be argued that if they are regularly valued then they already include any effects of CDS spread changes (at least for Goodwill).  In as much as an Accounting impairment to Goodwill is recognized this will be the case.  However, Accounting impairments require a particular standard of evidence \cite{CLW08} whereas CDS curves are simply observed day-to-day.  Thus CVA on Goodwill from CDS spread changes is relevant.  Additionally, if Goodwill behaves as a stock, i.e. has risk neutral rate of return, then we have shown its CVA is 100\%\ of its current value.   At least in the example considered for the LCFI there is no clear evidence that the CDS spread widening was included in the Goodwill as its value hardly changed between 2008YE and 2009Q1. In the future Accounting and Regulatory bodies may clarify this point.

We have helped to complete the picture of CVA and liquidity by considering unbooked positions and trades, i.e. trades that are required by the trades already in present for their funding, specifically collateralized trades, and (at least) assets explicitly reported in financial statements.  As well as Goodwill, other firm-level assets and balance sheet item such as Equity should also be included for firm-level CVA reporting.

Note that although we have used public information about a LCFI in this paper no conclusions about any LCFI  should be made or inferred.  We only used sufficient data for some partial examples, in no way does this paper attempt any commentary on the LCFI in question.

We acknowledge that trades and assets not in position-keeping systems may be difficult for accounting and regulatory bodies to include in their normal frameworks but leaving them out has produced a potentially unbalanced picture of financial institution PnL. Just how unbalanced depends on going into detail for each specific institution.  This is an appropriate topic for future research, accounting institutions, and regulatory bodies --- but beyond the scope of the present paper.

\section*{Appendix: Curves and Surfaces}

\begin{figure}[htb]
\begin{center}
\includegraphics[width=0.6\textwidth,  clip=true, trim=0 0 0 0 ]{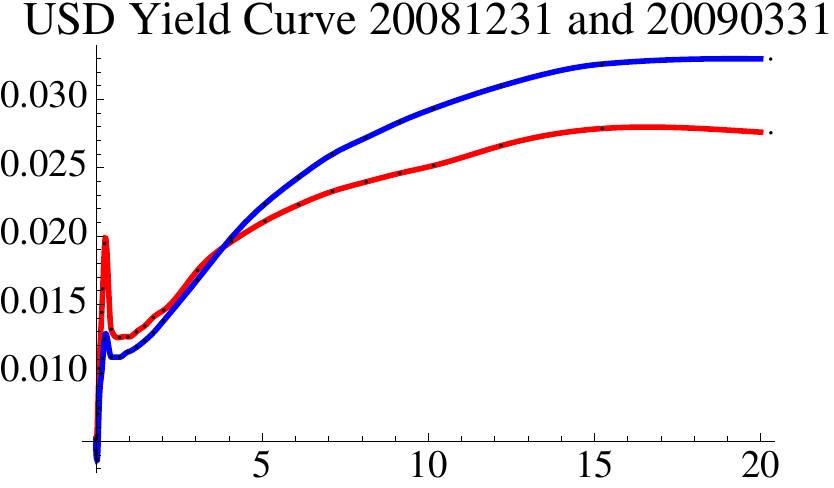}
\caption{USD Yield curves, the upper curve is for the earlier date.}
\end{center}
\end{figure}

\begin{figure}[htb]
\begin{center}
\includegraphics[width=0.6\textwidth,  clip=true, trim=0 0 0 0 ]{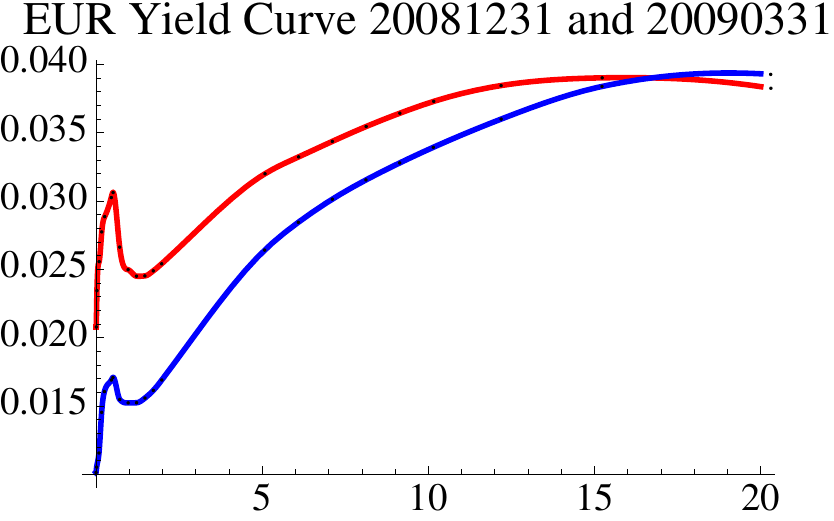}
\caption{EUR Yield curves, the lower curve is for the earlier date.}
\end{center}
\end{figure}

\begin{figure}[htb]
\begin{center}
\includegraphics[width=0.6\textwidth,  clip=true, trim=0 0 0 0 ]{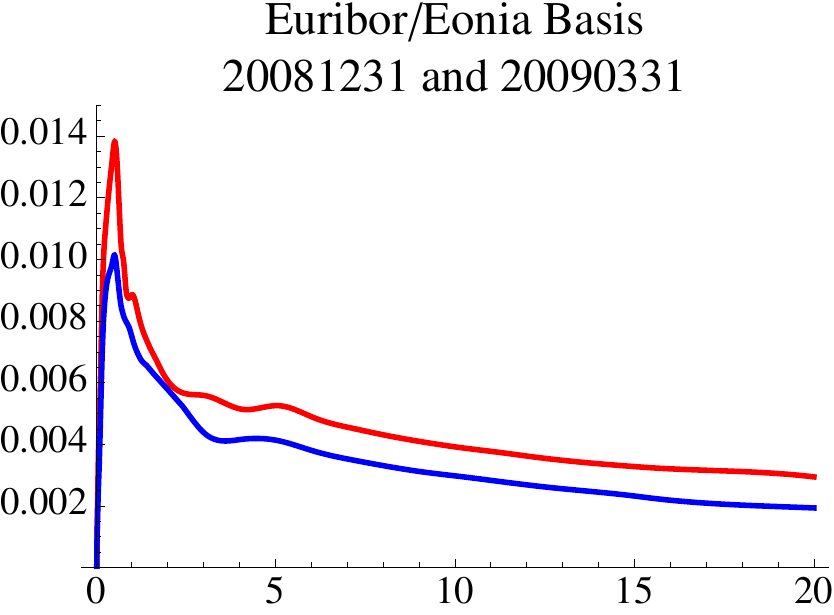}
\caption{Euribor/Eonia basis.  The lower curve is for the earlier date.}
\end{center}
\end{figure}

\begin{figure}[htb]
\begin{center}
\includegraphics[width=0.7\textwidth,  clip=true, trim=0 0 0 0 ]{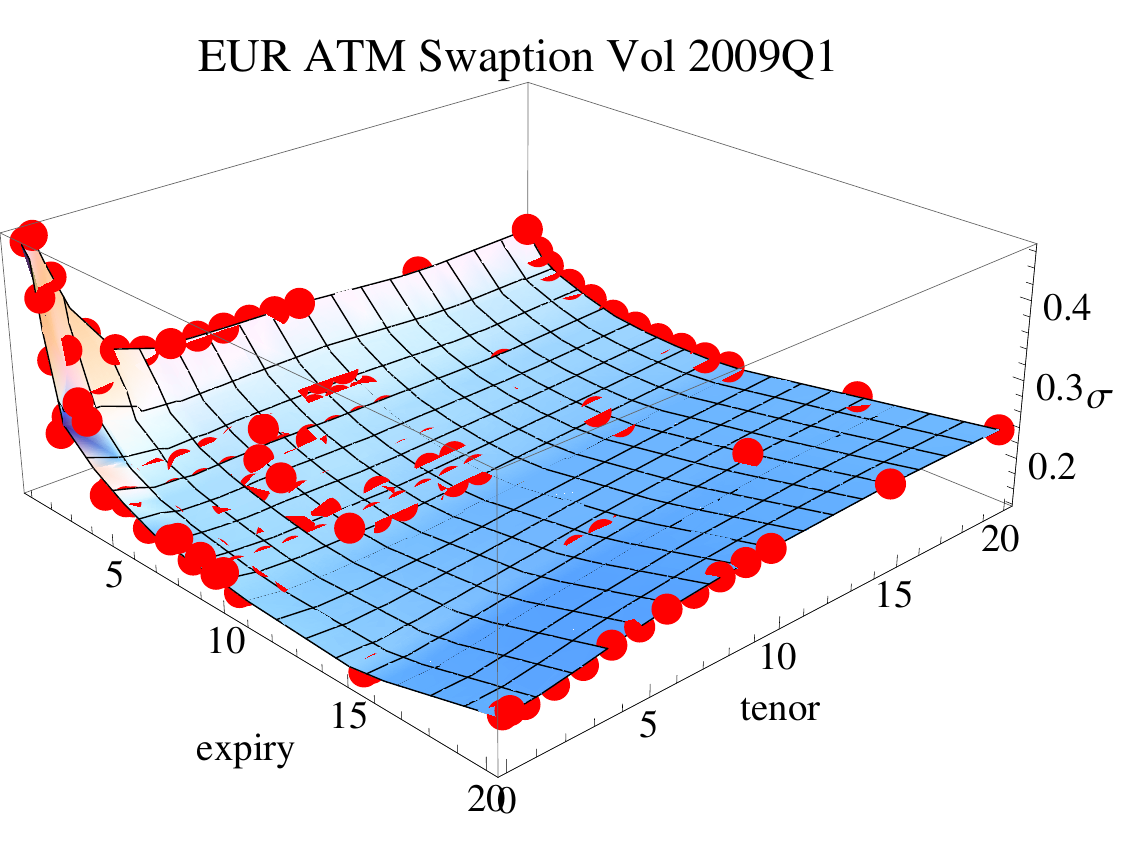}
\caption{EUR ATM Swaption implied volatility 2009Q1 and data points, the earlier date is similar but a little higher.}
\end{center}
\end{figure}

Yield curves, basis curves, and swaption volatilities.  N.B. we used swaption smile volatilities as well as the ATM surface presented below.

\bibliographystyle{alpha}
\bibliography{zotero}

\end{document}